\documentclass{jpp}

\usepackage{graphicx}
\usepackage{natbib}
\usepackage{amsmath}
\usepackage{color}

\ifCUPmtlplainloaded \else
  \checkfont{eurm10}
  \iffontfound
    \IfFileExists{upmath.sty}
      {\typeout{^^JFound AMS Euler Roman fonts on the system,
                   using the 'upmath' package.^^J}%
       \usepackage{upmath}}
      {\typeout{^^JFound AMS Euler Roman fonts on the system, but you
                   dont seem to have the}%
       \typeout{'upmath' package installed. JPP.cls can take advantage
                 of these fonts, if you use 'upmath' package.^^J}%
      }
  \else
  \fi
\fi

\ifCUPmtlplainloaded \else
  \checkfont{msam10}
  \iffontfound
    \IfFileExists{amssymb.sty}
      {\typeout{^^JFound AMS Symbol fonts on the system, using the
                'amssymb' package.^^J}%
       \usepackage{amssymb}%

      }{}
  \fi
\fi

\ifCUPmtlplainloaded \else
  \IfFileExists{amsbsy.sty}
    {\typeout{^^JFound the 'amsbsy' package on the system, using it.^^J}%
     \usepackage{amsbsy}}
    {\providecommand\boldsymbol[1]{\mbox{\boldmath $##1$}}}
\fi

\newcommand{\bskp}{\boldsymbol{k_\perp}}
\newcommand{\bsk}{\boldsymbol{k}}

\newcommand{\bfo}{\overline{\beta_{\rm f}}}
\newcommand{\gfo}{\overline{\gamma_{\rm f}}}

\title[Corrugation of a pulsar wind termination shock]{A corrugated termination shock in pulsar wind nebulae?}

\author[M. Lemoine]%
{M\ls A\ls R\ls T\ls I\ls N\ns L\ls E\ls M\ls O\ls I\ls N\ls E%
  \thanks{Email address for correspondence: lemoine@iap.fr}}

\affiliation{Institut d'Astrophysique de Paris, CNRS - UPMC, 98
  bis boulevard Arago, F-75014 Paris, France
}

\begin{document}

\maketitle

\begin{abstract}
Successful phenomenological models of pulsar wind nebulae assume
efficient dissipation of the Poynting flux of the magnetized
electron-positron wind as well as efficient acceleration of the pairs
in the vicinity of the termination shock, but how this is realized is
not yet well understood. The present paper suggests that the
corrugation of the termination shock, at the onset of non-linearity,
may lead towards the desired phenomenology. Non-linear corrugation of
the termination shock would convert a fraction of order unity of the
incoming ordered magnetic field into downstream turbulence, slowing
down the flow to sub-relativistic velocities. The dissipation of
turbulence would further preheat the pair population on short length
scales, close to equipartition with the magnetic field, thereby
reducing the initial high magnetization to values of order
unity. Furthermore, it is speculated that the turbulence generated by
the corrugation pattern may sustain a relativistic Fermi process,
accelerating particles close to the radiation reaction limit, as
observed in the Crab nebula. The required corrugation could be induced
by the fast magnetosonic modes of downstream nebular turbulence; but
it could also be produced by upstream turbulence, either carried by
the wind or seeded in the precursor by the accelerated particles
themselves.
\end{abstract}


\section{Introduction}\label{sec:intr}
Pulsar wind nebulae (PWNe) have long been recognized as outstanding
laboratories of astro-plasma physics in extreme conditions, see
e.g. \citet{2009ASSL..357..421K} and \cite{2012SSRv..173..341A}, and
the Crab nebula, as a result of its proximity, plays a very special
role among those objects.

At the price of non-trivial assumptions regarding the physical
conditions behind the termination shock of the pulsar wind, which
separates the free streaming wind from the hot shocked wind in the
nebula, phenomenological models have been very successful in
explaining the general spectral energy distribution and morphological
features of the Crab nebula, using analytical calculations
\citep[e.g.][]{1984ApJ...283..694K,1984ApJ...283..710K,1996MNRAS.278..525A}
or increasingly sophisticated numerical simulations
\citep[e.g.][]{2003MNRAS.344L..93K,2004MNRAS.349..779K,2003A&A...405..617B,2004A&A...421.1063D,2014MNRAS.438..278P},
see also \citet{2015arXiv150302402A} and \citet{2015SSRv..191..391K}
for reviews.

Nonetheless, various puzzles plague the current understanding of the
microphysics of PWNe; among them, two are particularly noteworthy: how
the wind converts its Poynting flux -- which supposedly far dominates
the particle kinetic energy at the base of the wind -- into particle
thermal energy behind the termination shock, reaching rough
equipartition between these components; and how particle acceleration
takes place behind the termination shock.

The present paper examines a speculative scenario, which could
potentially solve part of the above puzzles; it specifically assumes
that the termination shock of the pulsar wind is non-linearly
corrugated, the precise meaning of this being given in
Sec.~\ref{sec:def}. It then shows that such corrugation efficiently
converts an incoming ordered magnetic energy into turbulence, thereby
slowing down appreciably the flow velocity behind the termination
shock. A significant part of the turbulence can be further dissipated
through collisionless effects on short length scales, leading to
pre-acceleration of the pairs, up to close equipartition with the
incoming magnetic energy. The corrugation of the shock may thus
achieve efficient dissipation of the incoming Poynting flux, in a way
that is reminiscent of the dissipation through reconnection of a
striped wind in the equatorial plane \citep{2003MNRAS.345..153L}.
Finally, it is speculated (and argued) that the turbulence seeded by
corrugation may also sustain an efficient Fermi process, leading to a
particle spectrum close to what is observed.

This paper is organized as follows: Sec.~\ref{sec:corr} discusses the
physics of a corrugated shock wave in the MHD limit;
Sec.~\ref{sec:acc} recalls some results on the collisionless damping
of relativistic MHD waves in a relativistic plasma, then discusses the
physics of particle pre-acceleration in the resulting turbulence and
the development of a relativistic Fermi process at high
energies. Sec.~\ref{sec:disc} discusses various possible sources of
corrugation and examines how the present results can be applied to
PWNe. Finally, Sec.~\ref{sec:conc} provides a summary of the
discussion and some conclusions.

\section{A relativistic corrugated termination shock}\label{sec:corr}
\subsection{Definitions}\label{sec:def}
Assume that the termination of the pulsar wind, which separates the
cold magnetized incoming wind, from the hot shocked wind, is
corrugated. Potential sources of corrugation will be addressed in
Sec.~\ref{sec:disc}. For simplicity, neglect effects of spherical
symmetry, which are not important to the present discussion, and
therefore consider a planar shock, moving at velocity $\bfo$ along the
$x-$direction relative to the downstream plasma, i.e. relative to the
nebula. The (uncorrugated) shock surface is defined by
\begin{equation}
\overline\Phi(x)\,=\,x-\bfo c t\,=\,0
\end{equation}
with corresponding shock normal $4-$vector:
\begin{equation}
\overline\ell_\mu\,=\,\frac{\partial_\mu\overline\Phi}{\left\vert\partial_\alpha\overline\Phi\partial^\alpha\overline\Phi\right\vert^{1/2}}\,=\,
\left(-\gfo\bfo,\gfo,0,0\right)\label{eq:ell0}
\end{equation}
where $\gfo\,\equiv\,\left(1-\bfo^2\right)^{-1/2}$
denotes the bulk Lorentz factor of the shock front relative to
downstream. 

Corrugation can be described by a perturbation of the shock surface:
\begin{equation}
  \Phi(x) \,=\,\overline\Phi(x) - \delta X\left(\boldsymbol{x_\perp},t\right)\,=\,0
  \label{eq:dphidX}
\end{equation}
where $\boldsymbol{x_\perp}\,\equiv\,\left(y,z\right)$ represents the
coordinates in the (uncorrugated) shock front plane. For simplicity,
consider for the purpose of this subsection a corrugation on a single
length scale characterized by a wavenumber $\bskp\,=\,(k_y,k_z)$
defined in the uncorrugated shock front plane, with harmonic behavior
at frequency $\omega_{\bsk}$:
\begin{equation}
  \delta X\left(\boldsymbol{x_\perp},t\right)\,=\,\delta X_{\bsk}\,e^{-i\omega_{\bsk}t + i\bskp\cdot\boldsymbol{x_\perp}}
\end{equation}
The frequency $\omega_{\bsk}$ is directly related to
$\bsk\,=\,(k_x,\bskp)$ and to the nature of the wave inducing the
corrugation of the shock front~\citep{LRG16}; $k_x$ represents here
the $x-$wavenumber of that mode. At a relativistic shock, one
typically has $\omega_{\bsk}\,\sim\,\left\vert\bsk\right\vert$, hence
this scaling is retained in the following.

In the linear approximation, the first order perturbation of the shock normal is written:
\begin{equation}
\delta\ell_\mu\,=\,-\frac{\partial_\mu\delta X}{\left\vert
\partial_\alpha\overline\Phi\partial^\alpha\overline\Phi\right\vert^{1/2}}
+ \frac{\partial_\mu\overline\Phi\,\partial_\beta\delta X\partial^\beta\overline\Phi}{\left\vert
\partial_\alpha\overline\Phi\partial^\alpha\overline\Phi\right\vert^{3/2}}
\end{equation}
or, for the above single wave corrugation pattern,
\begin{equation}
\delta\ell_{\bsk\,\mu}\,=\,\biggl(i\gfo^3\,\omega_{\bsk}\delta X_{\bsk}/c,\,\,
-i\gfo^3\bfo\,\omega_{\bsk}\delta X_{\bsk}/c,\,\, -ik_y\,\gfo\delta X_{\bsk},
-ik_z\,\gfo\delta X_{\bsk}\biggr)
\end{equation}
A strongly corrugated shock is such that the above linear
approximation breaks down, i.e. $\left\vert\delta \ell\right\vert
\,\gtrsim\,1$, which translates into
\begin{eqnarray}
  \gfo\left\vert k_\perp\, \delta X_{\bsk}\right\vert &\,\gtrsim\,&1\nonumber\\
  \gfo^2\left\vert \omega_{\bsk}\, \delta X_{\bsk}\right\vert&\,\gtrsim\,&1
\end{eqnarray}
Both conditions express the fact that the departure from the
unperturbed shock surface becomes larger than the perpendicular
wavelength \emph{in the shock front rest frame}. For a moderately
magnetized shock wave, with $\sigma_1\,\lesssim\,3$, one has
$\gfo\,\sim\,1$, hence a shock front at the onset of non-linear
corrugation is such that $\left\vert k\, \delta X\right\vert\,\sim\,1$
(with $\delta X$ expressed in the downstream rest frame as
previously).

If the incoming upstream flow is strongly magnetized,
i.e. $\sigma_1\,\gtrsim\,3$, $\gfo$ becomes larger than
unity. However, the mean shock velocity $\beta_{\rm f}$ along the
shock normal can be strongly modified by corrugation; as discussed in
the following, in particular, the generation of turbulence in the
non-linearly corrugated shock transition can reduce the actual
$\beta_{\rm f}$ to sub-relativistic values, implying $\gamma_{\rm
  f}\,\sim\,1$. It is therefore speculated that $\left\vert k\, \delta
X\right\vert\,\sim\,1$ remains a valid threshold for non-linear
corrugation in the large magnetization regime
$\sigma_1\,\gtrsim\,3$.

Acutally, one could potentially envisage even larger values of $\delta
X$; however, non-linear back-reaction would likely limit $\delta X$ to
the above threshold of non-linearity, hence this value is retained in
the following.

Interestingly, the condition $\left\vert k\,\delta X_{\bsk}\right\vert
\,\gtrsim\,1$ is compatible with a very small amplitude $\vert\delta
X\vert$, as measured relative to the scale of the termination shock
(noted $R$), provided $\left\vert k\, R\right\vert \,\gg\,1$. In other
words, the shock may be strongly corrugated on short spatial scales
which are not observable at large distances, but which remain
macroscopic compared to the thickness of the actual shock transition.
The latter requirement is not essential for the present scenario, but
the MHD description that this model uses can only be applied on scales
much larger than the shock thickness, which is set by kinetic
physics. To provide quantitative estimates, consider the case of the
Crab nebula: if electrons are inflowing through the shock with a
Lorentz factor $\gamma_{\rm w}=10^4\gamma_{\rm w,4}$ (relative to the
downstream-nebula rest frame), the typical gyroradius of these
particles in the downstream magnetic field (strength $B_{\rm
  d}\,\sim\,0.1\,$mG), $r_{\rm g}\,\sim\,10^{11}\gamma_{\rm w,4}\,$cm
sets the typical thickness of the shock transition layer. Therefore,
corrugation is envisaged here on all scales $k^{-1}$ larger than the
above and smaller than $R\,\sim\,3\times 10^{17}\,$cm. As discussed in
Section~\ref{sec:disc}, this range of scales encompasses the gyroradii
of all accelerated particles, even the highest energy ones, indicating
that corrugation may exert a strong influence on their kinematics.

\subsection{A particular non-linear solution}
In the non-linear regime where the above perturbative description
breaks down, it is possible to extract an analytical solution of the
shock crossing equations in a 2D limit, in which all quantities remain
unperturbed along the direction of the background magnetic field
(taken to be $\boldsymbol{z}$ here), see \citet{LRG16}. The downstream
quantities characterizing the state of the plasma can then be
expressed at any time, on the corrugated shock front, in terms of the
shock normal. The magnitude of the shock corrugation amplitude, which
depends on the past history of the accumulated flow, is related
through a non-trivial differential relation to the shock normal, see
Eqs.~(\ref{eq:ell0}) and (\ref{eq:dphidX}) in particular. The present
description does not attempt to describe this corrugation amplitude
but to describe the state of the downstream plasma on the shock front.

\begin{figure}
\centerline{\includegraphics[width=0.8\columnwidth]{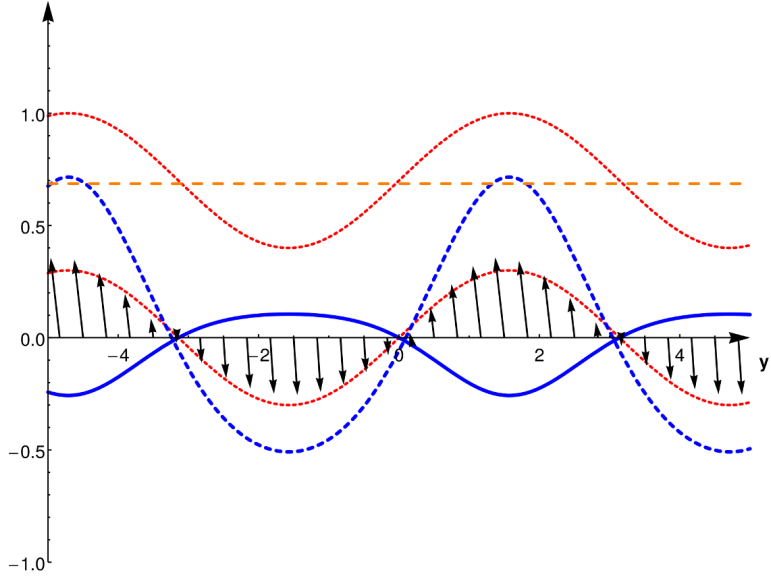}}
\caption{Various quantities plotted at $t\,=\,0$ as a function of $y$,
  the coordinate along the direction $\boldsymbol{y}$, which is both
  perpendicular to the shock normal ($\boldsymbol{x}$) and to the
  direction of the background magnetic field ($\boldsymbol{z}$), in a
  case of non-linear corrugation described by Eq.~(\ref{eq:nlex}),
  assuming no rippling along $\boldsymbol{z}$. Upper sinusoidal dotted
  line: $-\ell_0/\ell_1$ describing the spatial behavior of the
  temporal component of the shock normal; the average value over $y$,
  i.e. $0.7$, represents the average velocity of the shock front
  relative to downstream. Lower sinusoidal dotted line:
  $-\ell_2/\ell_1$, describing the rippling of the shock front in the
  $\boldsymbol{y}$ direction. Horizontal dashed curve: (relativistic)
  Alfv\'en $3-$velocity of waves in the downstream plasma, whose
  modulations are at too small amplitude to emerge on this figure. Solid
  (blue) line: $y-$component of the downstream flow $3-$velocity (on
  the shock front). Dashed (blue) thick line: $x-$component of the
  downstream flow $3-$velocity. Finally, the arrows indicate the
  direction of the downstream $3-$velocity in the $(y,x)$ plane (for
  this arrow representation, the ordinate axis should be understood as
  indicating the $x-$direction, while the abscissa points into the
  $y-$direction).
  \label{fig:nlcorr}}
\end{figure}

An example is shown in Fig.~\ref{fig:nlcorr}, which assumes an
upstream magnetization parameter $\sigma_1\,=\,1$, a relative Lorentz
factor between up- and down-stream $\gamma_1\,=\,100$, and a shock
normal $4-$vector arbitrarily set to:
\begin{equation}
  \ell_\mu\,=\,\left\{-0.98\,+\, 0.42\sin\left(t-y\right),\,\,
  1.40,\,\, 0.42\sin\left(t-y\right),\,\,0\right\}\label{eq:nlex}
\end{equation}
In the downstream rest frame, this $4-$normal describes a mean shock
velocity of $\beta_{\rm f}\,\simeq\,0.7$ (corresponding to the
unperturbed shock crossing conditions for the above values of the
magnetization and flow velocity), with harmonic corrugation slightly
below the onset of the non-linear limit.

For the particular case of perturbations confined to the
$y-$direction, the equations of shock crossing lead to $B_{x\vert
  2}\,=\,B_{y\vert 2}\,=\,0$: the magnetic field retains only a
$z-$component, modulated along $\boldsymbol{y}$. Note that $B_{x\vert
  2}$ and $B_{y\vert 2}$ vanish on the corrugated shock surface, but
not necessarily further downstream. One would need to follow the
characteristics of the system to study the evolution of these two
quantities downstream of the shock; in the linear limit at least, one
can show that the outgoing wave modes develop a net $\delta B_{x\vert
  2}$ and $\delta B_{y\vert2}$ further downstream of the
shock. Moreover, in a more general case with $k_z\,\neq\,0$, both
components $\delta B_{x\vert 2}$ and $\delta B_{y\vert 2}$ would not
vanish on the shock surface.

As illustrated by Fig.~\ref{fig:nlcorr}, the shock crossing equations
imply the existence of significant velocity fluctuations along $x$ and
$y$, with different modulations. In turn, these generate sheared flows
with non-vanishing $\boldsymbol{\beta}\times\boldsymbol{B}$ in the
downstream, leading to the stretching and compression of magnetic
field lines. These flow motions also generate convective electric
fields, $\boldsymbol{E}\,=\,-\boldsymbol{\beta}\times\boldsymbol{B}$,
with $\vert\boldsymbol{E}\vert\,\sim\,\vert\boldsymbol{B}\vert$
because $\vert\beta\vert$ is close to unity. The resulting turbulence
should thus be prone to dissipation and particle acceleration on short
time scales.

\subsection{Jump conditions at a corrugated shock front}
As the upstream plasma inflows through the corrugated shock, the
ordered magnetic energy is thus converted in part into turbulence
modes. Assume that the MHD fluid immediately behind the shock can be
described by an enthalpy density $w_2$ and pressure $p_2$ with
equation of state $w_2\,\simeq\,4p_2$ (relativistically hot fluid),
and by a magnetic field
$\boldsymbol{B}\,=\,\boldsymbol{B_2}+\boldsymbol{\delta B_2}$, with
$\left\langle\boldsymbol{\delta B_2}\right\rangle\,=\,0$. The index
$_2$ applies to downstream quantities; upstream quantities will be
indexed with $_1$. The average can be taken in the ensemble of
realizations of the corrugation, or as usual, in the ergodic
hypothesis, along the shock front plane.

On spatial scales much larger than the thickness of the shock, but
much smaller than the corrugation amplitude $\vert\delta X\vert$, the
jump conditions at the shock are expressed through the integration of
the conservation laws along the perturbed normal direction. However,
the present discussion is rather concerned with computing the
asymptotic behavior of the downstream plasma, on a distance scale
$\,\gg\,\vert\delta X\vert$ away from the shock. On such scales, the
shock can be seen as a planar discontinuity, although the jump
conditions should reflect the fact that turbulence has been generated
in the transition layer. These jump conditions should thus be written
in the downstream plasma rest frame as
\begin{eqnarray}
  \left[n u^\mu \ell_\mu\right]&\,=\,&0\,\nonumber\\
  \left[T^{\mu\nu}\ell_\mu\right]&\,=\,&0\,
\end{eqnarray}
where the shock normal $\ell_\mu\,=\,\left(-\gamma_{\rm f}\beta_{\rm
  f},\gamma_{\rm f},0,0\right)$ as in Eq.~(\ref{eq:ell0}) above. A
distinction has to be made however in the notations: in
Eq.~(\ref{eq:ell0}), the overline symbols indicate that the solution
applies in the absence of corrugation, while in the present case,
corrugation is assumed to be present on small spatial scales, leading
to the production of downstream MHD turbulence, so that the value of
$\beta_{\rm f}$ which is determined further below differs from $\bfo$
above.

The energy-momentum tensor in the ideal MHD approximation is written
\begin{eqnarray}
T^{\mu\nu}&\,=\,&\left(w + \frac{b_\alpha b^\alpha}{4\pi}\right)u^\mu u^\nu\,+\,
\left(p+\frac{b_\alpha b^\alpha}{8\pi}\right)\eta^{\mu\nu}-\frac{b^\mu b^\nu}{4\pi}
\end{eqnarray}
in terms of the magnetic $4-$vector $b^{\mu}$:
\begin{equation}
b^\mu\,=\,\left[u^iB_i,\left(\boldsymbol{B}+u^iB_i\,\boldsymbol{u}\right)/u^0\right]
\end{equation}
where $u^\mu$ represents the turbulent fluid $4-$velocity behind the
shock, and latin indices represent spatial indices.

The average energy-momentum tensor of the turbulent downstream fluid
is given by the correlators $\left\langle
b^{\alpha}b^{\beta}\right\rangle$, which can be calculated in full
generality. In order to derive simple estimates however, the velocity
field and the magnetic perturbations are assumed uncorrelated, i.e.
\begin{equation}
  \left\langle u^\alpha u^\beta B_{2\,i} B_{2\,j}\right\rangle\,=\,\left\langle u^\alpha u^\beta\right\rangle\left\langle B_{2\,i} B_{2\,j}\right\rangle
\end{equation}
and correlators involving odd powers of the turbulent fluid
$3-$velocity are assumed to vanish; in particular, $\langle
u^i\rangle\,=\,0$ by definition of the downstream rest
frame. Finally, one may assume isotropic turbulence, meaning
\begin{equation}
  \left\langle \delta B_{2\,i}\delta B_{2\,j}\right\rangle\,=\,\frac{1}{3}\delta_{ij}\,
  \left\langle\delta B_2^2\right\rangle
\end{equation}
This assumption allows to simplify the calculations, but it is not
crucial to the present analysis, as discussed further on.

Then one writes
\begin{equation}
  \left\langle b^{\alpha}b^{\beta}\right\rangle\,=\,\left\langle
  b_{(0)}^{\alpha}b_{(0)}^{\beta}\right\rangle+\left\langle
  b_{(1)}^{\alpha}b_{(1)}^{\beta}\right\rangle
\end{equation}
where $b_{(0)}^\alpha$ does not contain any $\delta B_i$ component,
while $b_{(1)}^\alpha$ does not contain any $B_i$ term. Assuming that
the average background field lies along $\boldsymbol{z}$, one readily
obtains
\begin{equation}
  \left\langle b_{(0)}^{\alpha}b_{(0)}^{\beta}\right\rangle\,=\,
  \delta^{\alpha}_0\delta^\beta_0\langle u_z^2\rangle B_2^2 +
  \delta^{\alpha}_i\delta^\beta_j\left[\left\langle\frac{1}{\gamma^2}\right\rangle\delta^{i\,3}\delta^{j\,3}
    +
    2\left\langle\frac{u_z^2}{\gamma^2}\right\rangle\delta^{i\,3}\delta^{j\,3}
    + \left\langle\frac{u^iu^ju_z^2}{\gamma^2}\right\rangle\right]B_2^2
\end{equation}
and
\begin{equation}
   \left\langle b_{(1)}^{\alpha}b_{(1)}^{\beta}\right\rangle\,=\,
   \delta^{\alpha}_0\delta^\beta_0\frac{1}{3}\langle u^2\rangle \left\langle\delta B_2^2\right\rangle +
   \delta^{\alpha}_i\delta^\beta_j\frac{1}{3}\left[\left\langle\frac{1}{\gamma^2}\right\rangle\delta^{i\,j}
     + 2\left\langle\frac{u^iu^j}{\gamma^2}\right\rangle + \left\langle
     \frac{u^iu^ju^2}{\gamma^2}\right\rangle\right]\left\langle\delta B_2^2\right\rangle
\end{equation}
A key observation here is that these correlators do not share the same
dependence. To proceed further and obtain a more tractable expression
leading to a simple estimate of the modified jump conditions, assume
that the turbulence in the downstream rest frame is moderately
relativistic; this is actually ensured if the corrugation is mildly
non-linear and $\sigma_1$ not large compared to unity, as discussed in
\citet{LRG16}. The following expressions thus retain only the leading
order terms in powers of $u$ in the above correlators, which then
reduce to standard non-relativistic expressions for magnetized
turbulence:
\begin{eqnarray}
  \left\langle b_{(0)}^{\alpha}b_{(0)}^{\beta}\right\rangle&\,\approx\,&
  \delta^{\alpha}_3\delta^\beta_3B_2^2\nonumber\\
  \left\langle b_{(1)}^{\alpha}b_{(1)}^{\beta}\right\rangle&\,\approx\,&
   \delta^{\alpha}_i\delta^\beta_j\delta^{ij}\frac{1}{3}\left\langle\delta B_2^2\right\rangle
\end{eqnarray}
It should be understood that this set of approximations is intended to
show in a quantitative way how corrugation affects the jump conditions
at the MHD shock. Qualitatively speaking however, the key point is
that the correlators of the turbulence modes contained in
$\left\langle b^{\alpha}b^{\beta}\right\rangle$ differ from those of
the average background field.

With the above approximations, the non-zero downstream energy-momentum
tensor components can be written:
\begin{eqnarray}
T_2^{00}&\,\approx\,&W_2-P_2\,\nonumber\\
T_2^{11}&\,\approx\,&P_2-\frac{1}{3}\frac{\left\langle\delta B_2^2\right\rangle}{4\pi}\,\nonumber\\
T_2^{22}&\,\approx\,&P_2-\frac{1}{3}\frac{\left\langle\delta B_2^2\right\rangle}{4\pi}\,\nonumber\\
T_2^{33}&\,\approx\,&P_2-\frac{B_2^2}{4\pi}-\frac{1}{3}\frac{\left\langle\delta B_2^2\right\rangle}{4\pi}\ ,
\end{eqnarray}
with $W_2\,\equiv\, w_2 + B_2^2/(4\pi) + \left\langle\delta
B_2^2\right\rangle/(4\pi)$ a generalized enthalpy density and
$P_2\,\equiv\, p_2 + B_2^2/(8\pi) + \left\langle\delta
B_2^2\right\rangle/(8\pi)$ a generalized pressure.

The shock jump conditions for energy and momentum fluxes are then expressed as
\begin{eqnarray}
  -\beta_{\rm f}\left(W_2-P_2\right)&\,=\,&\left(\beta_1-\beta_{\rm
    f}\right)\gamma_1^2 W_1 + \beta_{\rm f}P_1 \nonumber\\
  P_2-
  \frac{1}{3}\frac{\left\langle\delta B_2^2\right\rangle}{4\pi}&\,=\,&\left(\beta_1-\beta_{\rm f}\right)\gamma_1^2\beta_1 W_1
  + P_1
\end{eqnarray}
The above equations are written in the downstream rest frame, so that
$\beta_1$ corresponds to the velocity of upstream relative to
downstream. Neglecting $P_1$ in front of $P_2$ (or, alternatively
$\gamma_1^2 W_1$) and assuming a hot relativistic plasma downstream,
$p_2\,=\,w_2/4$, one obtains easily
\begin{equation}
  \beta_1\beta_{\rm f}\,=\,-\frac{\displaystyle{w_2 + \frac{B_2^2}{2\pi} +\frac{\left\langle\delta B_2^2\right\rangle}{6\pi}}}{\displaystyle{3w_2 + \frac{B_2^2}{2\pi} + \frac{\left\langle\delta B_2^2\right\rangle}{2\pi}}}\ .
\end{equation}
This equation reduces to the standard result $\beta_{\rm
  f}\,\rightarrow\,1/3$ in the limits $\beta_1\,\rightarrow\,-1$
(ultra-relativistic limit), $B_2\,\rightarrow\,0$ and
$\left\langle\delta B_2^2\right\rangle^{1/2}\,\rightarrow\,0$
(hydrodynamic shock), e.g.~\citet{1999JPhG...25R.163K}. In the highly
magnetized and uncorrugated case, meaning $w_2\,\ll\,B_2^2/(4\pi)$ and
$\left\langle\delta B_2^2\right\rangle^{1/2}\,\rightarrow\,0$, one
also recovers $\beta_1\beta_{\rm f}\,\simeq\,-1$, indicating that the
shock moves away from downstream at a relativistic velocity. As
discussed by \citet{1984ApJ...283..694K}, the mismatch between this
solution and the general morphology of the Crab nebula points towards
a smaller than unity magnetization parameter behind the termination
shock.

More interestingly, if $\left\langle\delta B_2^2\right\rangle^{1/2}$
is not negligible compared to $B_2$, one finds a solution with a shock
moving away from downstream at sub-relativistic velocities,
independently of how strongly magnetized the flow initially
was. Consider for instance the case of equipartition
$\left\langle\delta B_2^2\right\rangle^{1/2}\,\sim\,B_2$ and
$w_2\,\ll\,B_2^2/(4\pi)$, which is effectively what one expects if
corrugation is midly non-linear. Then $\beta_{\rm f}\,\simeq\,2/3$ for
$\beta_1\,\simeq\,-1$. Other interesting limits are
$\left\langle\delta B_2^2\right\rangle\,\gg\,B_2^2$ (strong
corrugation), leading to $\beta_{\rm f}\,\simeq\,1/3$ as in a pure
hydrodynamical shock; or $\left\langle\delta
B_2^2\right\rangle\,\sim\,B_2^2\,\sim\,4\pi w_2$, leading to
$\beta_{\rm f}\,\simeq\,0.5$.

These particular solutions emerge whenever the field line tension of
the turbulence can contribute to the $xx$ component of $T^{\mu\nu}$,
i.e. whenever the correlators $\left\langle b^\mu b^\nu\right\rangle$
possess a non-trivial $xx$ component. As discussed here and in the
previous Section, such a component could arise from the shock
corrugation directly or from the non-linear processing of the magnetic
modulations induced by shock corrugation. The following sub-sections
also argue that the partial dissipation of such turbulence would
preheat the pairs and thus lead to a solution with a moderate shock
velocity relative to downstream (corresponding to the last case of
equipartition discussed above).

\section{Particle pre-acceleration and acceleration}\label{sec:acc}

\subsection{Collisionless damping of hydromagnetic waves}\label{sec:damp}
Even if $w_2\,\ll\, B_2^2/(4\pi),\,\, \left\langle\delta B_2^2\right\rangle/(4\pi)$ immediately
downstream of the shock, various dissipative effects will transfer
energy from the magnetized turbulence to the particles, thereby
decreasing the magnetization of the plasma to values of order unity
(see below). The present Section discusses the collisionless damping
of magnetosonic modes at the Landau resonance; stochastic particle
acceleration will be discussed in Sec.~\ref{sec:dissip}; other
processes, such as turbulent reconnection, are of course plausible
sources of dissipation.

The collisionless damping of hydromagnetic waves in a relativistic
plasma has been discussed by \citet{1973ApJ...184..251B}.  At the
Landau resonance, they find
\begin{equation}
  \Im\omega\,\simeq\,\frac{\pi}{8}\Re\omega\,\frac{\delta B_k^2}{4\pi W_{\delta
      B_k}}\, \sin^2\theta\, \vert w_{\rm r}\vert\left(1-w_{\rm
    r}^2\right)^2\Theta\left(1-w_{\rm r}\right)\sigma^{-1}
\end{equation}
where $\omega$ corresponds to the wave frequency, $W_{\rm \delta
  B_k}\,\sim\,\delta B_k^2/(4\pi)$ to the wave energy density,
$\theta$ to the angle between $\boldsymbol{k}$ and $\boldsymbol{B_2}$,
$\Theta$ to the Heaviside function, $\sigma$ to the magnetization,
i.e. the ratio between the magnetic energy density and the electron
energy density. Finally,
\begin{equation}
  w_{\rm r}\,\equiv\,\frac{\Re\omega}{k c\cos\theta}
\end{equation}
is the resonance parameter.

As discussed in \citet{1973ApJ...184..251B}, the presence of the
Heaviside function limits the damping coefficient of waves by defining
a critical angle beyond which Landau damping vanishes. The real
frequency of magnetosonic waves propagating at an angle $\theta$ to
the magnetic field is given by
\begin{equation}
  \Re\omega\,=\,\frac{kc}{\sqrt{2}}\left\{\beta_+^2 + \beta_{\rm
    A}^2c_{\rm s}^2\cos^2\theta \pm \left[\left(\beta_+^2 + \beta_{\rm
      A}^2c_{\rm s}^2\cos^2\theta\right)^2-4\beta_{\rm A}^2c_{\rm s}^2\cos^2\theta\right]^{1/2}\right\}^{1/2}
\end{equation}
the plus sign pertaining to fast magnetosonic waves, and the minus
sign to slow magnetosonic waves; $\beta_+^2\,\equiv\,\beta_{\rm A}^2 +
c_{\rm s}^2-\beta_{\rm A}^2c_{\rm s}^2$ in terms of the (relativistic)
Alfv\'en velocity $\beta_{\rm A}$ and sound velocity $c_{\rm
  s}$. Assuming an isotropic bath of waves, one can calculate the
average damping coefficient, relatively to the mode wavenumber, as
\begin{equation}
  \frac{\left\langle\Im\omega\right\rangle}{kc}\,\simeq\,
  \frac{\pi}{16}\sigma^{-1}\int {\rm d}\theta\, \sin^3\theta\cos\theta\, w_{\rm r}^2\left(1-
  w_{\rm r}^2\right)^2\Theta\left(1-w_{\rm r}\right)
\end{equation}
For fast magnetosonic waves in a strongly magnetized plasma
($\sigma\,\gg\,1$), the fast magnetosonic wave phase velocity
approaches unity, so that the critical angle $\theta_{\rm
  c}\,\sim\,0$, implying
$\left\langle\Im\omega/kc\right\rangle\,\approx\,0$; those waves are
effectively undamped in the highly magnetized regime. At more moderate
magnetization, damping may become appreciable however: at
$\sigma\,=\,1$ for instance,
$\left\langle\Im\omega/kc\right\rangle\,\approx\, 10^{-4}$.

For slow magnetosonic waves, however, one finds typically
\begin{equation}
  \left\langle\Im\omega\right\rangle\,\simeq\,10^{-2}\,kc\,\sigma^{-1}
\end{equation}
at $\sigma\,\gtrsim\,1$.

The above should be considered as a lower limit to $\Im\omega$ since
the above neglects Landau-synchrotron damping effects, as well as
other dissipative effects. It nevertheless allows to set an upper
limit on the damping length associated the shock corrugation:
$\lambda_{\rm diss.}\,\simeq\,\beta_{\rm
  f}c\left\langle\Im\omega\right\rangle^{-1}$.

In the more realistic case of corrugation spread over a broad range of
$k-$modes, the above indicates that the short scale modes (large $k$)
will dissipate on short length scales $\propto k^{-1}$. Turbulence on
the large scales may dissipate through cascading to shorter scales,
followed by damping; given that fluctuations are mildly relativistic
behind the shock front if corrugation is midly non-relativistic, as
discussed above, the typical timescale of eddy cascading may be not
much larger than $(kc)^{-1}$, implying an efficient damping of slow
magnetosonic turbulence. The general picture that characterizes
non-linear shock corrugation is thus the generation of a turbulent
layer behind the shock front, a part of which dissipates on a small
length scale $\lambda_{\rm diss.}\,\sim\,\mathcal{O}(k_{\rm
  peak}^{-1})$, $k_{\rm peak}$ denoting the mode on which the maximum
turbulent power is concentrated.

At such a corrugated shock front, $\left\langle\delta B_2^2\right\rangle\,\sim\,B_2^2$ and slow
magnetosonic waves carry a significant fraction of the turbulence
magnetic energy density. Therefore, on a distance scale $\lambda_{\rm
  diss.}$, a fraction $\eta_{\rm s}$, with $\eta_{\rm s}$ not far
below unity, of the magnetic energy density has been dissipated into
particles, reducing the magnetization from
\begin{equation}
  \sigma_{2<}\,=\,\frac{\delta B_2^2 + B_2^2}{4\pi w_2}
\end{equation}
immediately downstream of the shock, to
\begin{equation}
  \sigma_{2>}\,=\,\frac{\left(1-\eta_{\rm s}\right)\sigma_{2<}}{1+\eta_{\rm s}\sigma_{2<}}
\end{equation}
beyond the dissipation layer.  Consequently, if
$\sigma_{2<}\,\gtrsim\,1$ and $\eta_{\rm s}\,\gtrsim\,1/\sigma_{2<}$,
the magnetization is reduced to values
$\sigma_{2>}\,\simeq\,(1-\eta_{\rm s})/\eta_{\rm s}$, of order unity,
independently of how high the magnetization initially was.

In this way, a magnetized relativistic corrugated shock wave can
dissipate efficiently the incoming magnetic energy into the shock,
leading to a near hydrodynamical shock with a magnetization of order
unity beyond $\lambda_{\rm diss.}$.

\subsection{Phenomenological model of particle pre-acceleration}\label{sec:dissip}
The acceleration of particles in a bath of magnetized turbulence can
be described in a phenomenological way through a Fokker-Planck
equation for the distribution function $f(\boldsymbol{p},t)$ of
particles\footnote{As discussed by \citet{1993PhyU...36.1020B},
  \citet{1996ApJ...461L..37B} and \citet{1999A&A...350..705P}, a
  rigorous model of particle acceleration in relativistic turbulence
  would require introducing more sophisticated kernels than the above
  Fokker-Planck operators. For mildly relativistic turbulence,
  however, the following Fokker-Planck analysis should provide a
  reasonable phenomenological model of particle acceleration.}:
\begin{equation}
  \frac{\partial }{\partial t}f \,=\,\frac{1}{p^2}\frac{\partial}{\partial p}\left(
  p^2 D_{pp}\,\frac{\partial}{\partial p}f\right) - \frac{1}{p^2}\frac{\partial}{\partial p}\left(\dot p p^2 f\right) - \frac{f}{\tau_{\rm esc}} + q(p)
\end{equation}
where $q$ models the injection of particles into the dissipation
region, $\tau_{\rm esc}$ the escape timescale out of the dissipation
region and $\dot p$ characterizes systematic energy
gains/losses. Dissipation is characterized by the diffusion
coefficient in momentum space
\begin{equation}
  D_{pp}\,=\,\frac{\left\langle \Delta p^2\right\rangle}{2\Delta t}
\end{equation}

The injection function takes the form $q(p)\,=\,\dot n/(4\pi p_0^2)\delta
(p-p_0)$, with $\dot n$ the density of particles per unit time
inflowing into the shock, as measured in the downstream rest frame;
the injection momentum $p_0$ is related to the shock Lorentz factor
through $p_0\,\simeq\,\gamma_1 mc$.

Ignoring systematic energy gains/losses and considering a stationary
state, standard techniques \citep[e.g.][]{1984A&A...136..227S} allow
to solve the above equation for various momentum dependences of
$D_{pp}$ and $\tau_{\rm esc}$. In the problem at hand, escape
presumably takes place through advection at velocity $\beta_{\rm f}$,
as the dissipation region is confined to a distance $\lambda_{\rm
  diss.}$ from the shock front, which itself moves away at velocity
$\beta_{\rm f}$. Thus $\tau_{\rm esc}\,\propto\, p^0$; as for the
momentum diffusion coefficient, it is written
\begin{equation}
  D_{pp}\,=\,\frac{p_0^2}{\tau_{\rm
      s,0}}\left(\frac{p}{p_0}\right)^{2+\alpha}
\end{equation}
where $\tau_{\rm s,0}$ characterizes the typical acceleration
timescale at momentum $p_0$. For $\alpha\,=\,0$, corresponding to the
simplest scaling with an interaction time in the turbulence that does
not depend on momentum, one finds
\begin{equation}
  f(p)\,=\,c_1\left(\frac{p}{p_0}\right)^{-\frac{3}{2}+\frac{1}{2}\sqrt{9+4\tau_{\rm s,0}/\tau_{\rm esc}}}
  \Theta(p_0-p)+
  c_2\left(\frac{p}{p_0}\right)^{-\frac{3}{2}-\frac{1}{2}\sqrt{9+4\tau_{\rm s,0}/\tau_{\rm esc}}}
  \Theta(p-p_0)
\end{equation}
where $c_1$ and $c_2$ are two integration constants related to $\dot
n$, $p_0$, $\tau_{\rm s,0}$ and $\tau_{\rm esc}$. The asymptotic
behavior at large momenta is thus a power-law
$f(p)\,\propto\,p^{-s_f}$ with index
\begin{equation}
  s_f\,=\,\frac{3}{2}+\frac{1}{2}\sqrt{9+4\tau_{\rm s,0}/\tau_{\rm esc}}
\end{equation}
which depends on how fast escape balances acceleration. One should
keep in mind that this index is that of the distribution function in
the acceleration zone, so that the index of the distribution function
per momentum interval ${\rm d}N/{\rm d}p\,\propto\,p^{-s}$ in this
acceleration zone is $s=s_f-2$. As to the distribution of escaping
particles, i.e. those that eventually populate the nebula, its index
is in principle modified by the escape rate, i.e.
\begin{equation}
\frac{{\rm d}N_{\rm esc}}{{\rm d}p}\,\propto \frac{1}{\tau_{\rm
    esc}}\frac{{\rm d}N}{{\rm d}p}
\end{equation}
but since $\tau_{\rm esc}\,\propto\,p^0$ here, the index remains
unchanged.

The above phenomenological model indicates that the dissipation of the
turbulence produced by corrugation leads to a power-law with index $s$
comprised between $1$ and $2$ if $\tau_{\rm s,0}\,\sim\,\tau_{\rm
  esc}$. The development of this power-law does not remain unbounded,
because most of the energy is then carried by particles of maximum
momentum, if $s<2$. Thus, in the absence of signicant energy losses,
one expects that the dissipation process saturates at a momentum such
that a significant fraction of the turbulence energy density has been
dissipated into the particles, i.e. such that the backreaction of the
dissipation process on the turbulence energy density cannot be
ignored. This takes place on a length scale $\lambda_{\rm diss.}$
which, accounting for all dissipative processes, characterizes the
width of the layer beyond which a fraction of order unity of the
magnetized turbulence energy has been dumped into the particles.

\subsection{Fermi acceleration}\label{sec:Fermi}
One may also expect the development of a relativistic Fermi process in
the above conditions. As a note of caution, however, the following
discussion remains qualitative and further work would be needed to
establish this statement on solid grounds.

At a relativistic shock of moderate to large magnetization, the Fermi
process is inhibited because of the superluminal nature of the
magnetic configuration. \citet{2006ApJ...645L.129L} and
\citet{2009MNRAS.393..587P} have discussed in some detail what
prevents Fermi cycles in such a configuration but it may be useful for
the present discussion to recall the salient points. Consider a planar
relativistic magnetized shock, with some turbulence upstream of the
shock, laid on a scale $\lambda$ assumed much larger than the typical
gyroradius $r_{\rm g}$ of accelerated particles in the downstream rest
frame. A key point is that the accelerated particles only probe a
region of size $r_{\rm g}$ during their Fermi cycles in such
turbulence: those that probe a deeper region downstream of the shock
are actually unable to return to the shock because of the superluminal
nature of the shock wave: in order to do so, particles would need to
diffuse across the magnetic field at an effective velocity larger than
$\beta_{\rm f}$. Thus, on the length scale $r_{\rm g}\,\ll\,\lambda$
probed by the particles, the turbulence appears as an essentially
coherent magnetic field. One may then show that incoming particles can
execute at most 1.5 Fermi cycles
up$\rightarrow$down$\rightarrow$up$\rightarrow$down in this
configuration before being advected downstream, away from the
shock. Those particles that are able to return once to the shock are
those whose momentum is oriented relative to the magnetic field in
such a way as to authorize a bounce on the downstream magnetic field,
pushing them back across the shock; but, for a near coherent upstream
magnetic field, this can happen only once for any particle.

For this reason, at a steady planar shock front, it has been proposed
that particle acceleration was associated with the development of
intense micro-turbulence on a scale $\,\ll\,r_{\rm g}$, in the shock
precursor~\citep{2006ApJ...645L.129L,2009MNRAS.393..587P}. This point
of view has been validated by particle-in-cell numerical simulations
which observe the concomitant development of micro-turbulence and of
particle
acceleration~\citep[e.g.][]{2008ApJ...682L...5S,2009ApJ...695L.189M,2013ApJ...771...54S}.

However, a crucial point of the previous argument is that the
direction of the coherent magnetic field line in the shock front plane
is conserved through the crossing of the shock. If this direction were
randomized through some process, then particles could bounce on the
downstream magnetic field with a non-zero probability at any Fermi
cycle and return to the shock. This bounce would be similar to an
isotropization of the particle directions downstream, i.e. similar to
a fast isotropic scattering process.  Therefore, it would lead to the
development of a Fermi process as modelled by early test-particle
Monte Carlo simulations (which implicitly assumed the non-conservation
of the direction of the magnetic field in the shock front plane)
\citep[e.g.][]{1998PhRvL..80.3911B,2000ApJ...542..235K,2001MNRAS.328..393A,2003ApJ...589L..73L},
with an index $s\,\simeq\, 2.2$ for ${\rm d}N/{\rm
  d}p\,\propto\,p^{-s}$.

Returning to the corrugated shock front, if $\left\langle\delta
B_2^2\right\rangle^{1/2} \,\sim \, B_2$ on a scale $k^{-1}$,
corrugation precisely does the above: even in the absence of upstream
turbulence, the in-plane direction of the downstream magnetic field is
randomized on a scale $k^{-1}$ by the generation of turbulence at the
corrugated shock. This should therefore lead to an efficient Fermi
process for particles of gyro-radius $r_{\rm g}\,\sim\,k^{-1}$,
although this should admittedly be explicitly demonstrated by
dedicated numerical simulations. Interestingly, if corrugation
sustains acceleration in the above way, the typical acceleration
timescale is then expected of order $r_{\rm g}/c$ in the shock frame,
as in the above test-particle simulations of the relativistic Fermi
process.

How this process affects particles of gyro-radius $r_{\rm
  g}\,\ll\,k^{-1}$ is not obvious. In a first approximation, one could
expect Fermi acceleration to be inoperant in that range of gyroradii
because those particles ``see'' the turbulent field as an essentially
coherent field. However, the scale $r_{\rm g}$ is then also much
smaller than the corrugation amplitude $\vert\delta X\vert$, hence the
time dependence of the corrugation, the rippled shock structure and
the relativistic turbulence existing in this layer could help sustain
acceleration. At the opposite extreme, $r_{\rm g}\,\gg\,k^{-1}$, the
turbulence may sustain acceleration, as long as the scattering
frequency $k^{-1}/r_{\rm g}^2$ remains larger than the advection
frequency $r_{\rm g,0}^{-1}$ in the background magnetic field (the
index 0 meaning that $r_{\rm g,0}$ is to be calculated relatively to
the background field).

In any case, one does not expect corrugation to occur on a single
scale $k$, but on a broad range of scales; in this case, the above
argument suggests that acceleration should at least take place for all
$r_{\rm g}$ in gyroresonance with the inertial range provided
$\left\langle\delta B_2^2\right\rangle^{1/2}\,\sim\,B_2$ can be
realized on those scales, i.e. provided corrugation is mildly
non-linear on all scales.

The overall picture becomes somewhat more complicated in the presence
of dissipation downstream of the shock, in particular how stochastic
energy gains interplay with systematic energy gains due to the Fermi
process. However, given that the Fermi process only transfers a small
fraction of the available energy to a small fraction of particles, one
may expect that dissipation would build up the hard power-law until
near equipartition with the magnetic field and that the Fermi
power-law would develop at higher momenta, until it saturates due to
energy losses.

\section{Discussion}\label{sec:disc}
The previous sections have argued that the corrugation of a magnetized
relativistic shock front, at the onset of the non-linear regime of
corrugation, could lead to interesting phenomenology. In particular,
it could provide an efficient source of dissipation of the magnetic
energy of the upstream flow, by converting the ordered magnetic field
into turbulence modes via its advection through the rippled shock,
with subsequent dissipation of the turbulent modes into supra-thermal
particle energy, reducing the initial magnetization to values of order
unity on a length scale $\lambda_{\rm diss.}$. It has also been shown
that this conversion into turbulence slows down appreciably the flow
velocity behind the shock (as now seen in the shock rest frame). The
present Section discusses possible sources of the corrugation and how
the above picture fits in a general model of pulsar wind nebulae.

\subsection{Sources of corrugation}
The stability of a shock front responding to small perturbations forms
a topic of research with a long history, going back to the pioneering
studies of ~\citet{1958JETP....6..739D} and
\citet{1958JETP....6.1179K}. Theorems guarantee the stability of
relativistic shock waves with a polytropic equation of state
\citep{1986PhFl...29.2847A,1987PhFl...30.1045A}, although a
corrugation instability may emerge in specific cases, as in e.g.  a
relativistic radiative shock ~\citep{TCST97}. In this limit of
instability to corrugation, small perturbations would induce a
deformation which would grow exponentially in time; this possibility
is however not considered in the present work.

Even if stable with respect to the corrugation instability, a shock
front may respond strongly to incoming perturbations, as discussed in
the above references, or in~\cite{LandauLifshitz87}. A recent
discussion of the response, possibly resonant, of a relativistic
magnetized shock to small amplitude disturbances can be found in
~\cite{LRG16}; its consequences are discussed further below.

Corrugation can be seeded by at least three sources: turbulence waves
originating from downstream and impacting the shock, turbulence modes
originating from upstream being advected through the shock, and
through instabilities seeded upstream of the shock by the accelerated
particles themselves.

As far as downstream waves are concerned, only fast magnetosonic modes
propagating at a group velocity larger than the shock velocity
$\beta_{\rm f}$ are able to induce corrugation. If $\delta
\psi_{\bsk}$ denotes the amplitude of the wave (with $\delta
\psi_{\bsk}\,\equiv\,\delta B_{\bsk}/B$), the corrugation amplitude
can be written to a reasonable accuracy as
\begin{equation}
  \left\vert\delta X_{\bsk}\right\vert\,\approx\,k^{-1}\left\vert\delta\psi_{\bsk}\right\vert
\end{equation}
The presence of $k^{-1}$ is directly related to the dimension of the
quantity of the l.h.s., of course. This result implies that non-linear
corrugation of the shock front requires non-linear fast magnetosonic
waves, meaning $\delta B/B\,\sim\,1$ for the incoming waves or, in
other words, that the turbulence carries an energy density comparable
to that of the average magnetic field advected through the shock. This
is certainly not a trivial requirement, but it seems to be fulfilled
at least in the numerical simulations of \citet{2009MNRAS.400.1241C}
which observed a strong backreaction of the nebular turbulence on the
shock.  One should also keep in mind that the above implicitly assumes
a stationary configuration; time-dependent turbulence might have a
stronger effect, as suggested by the discussion of
\citet{2012MNRAS.422.3118L}. Those simulations have not addressed the
dynamical range of scales over which the corrugation could be present;
presumably however, all scales up to the shock termination radius $R$
could be excited.

If turbulence is present upstream of the shock front to substantial
levels, as proposed recently \citep{2015arXiv151205426Z}, non-linear
corrugation should follow owing to the existence of a resonance of the
response of the shock to the incoming turbulence \citep{LRG16}. This
latter work has shown that the resonance occurs when the fast
magnetosonic mode, which is sourced downstream of the shock by the
shock corrugation, has a group velocity corresponding to $\beta_{\rm
  f}$, i.e. when this fast mode surfs along with the shock front. The
large response of the shock corrugation was then interpreted as the
build-up of fast magnetosonic energy on the shock front. Since the
group velocity is determined by the wavenumber $k_x$ at a given
$\bskp$, this resonance selects one value of the incoming $k_x$, with
typically $k_x\,\sim\,\vert\bskp\vert$ at a magnetized relativistic
shock where $\beta_{\rm f}$ does not lie far below unity.

Depending on $\bskp$, one may observe a formally infinite response of
the shock, or a large amplification of the incoming waves at the
resonance. This thus opens the possibility of reaching the threshold
of non-linear corrugation with a source whose energy content is less
than that of the incoming ordered magnetic field.  The study of
\citet{LRG16} has been conducted in linearized MHD, therefore it
cannot probe the deep non-linear regime of corrugation; it seems
reasonable to assume that, on those resonant scales, the shock is
corrugated with amplitude $\vert k_\perp \delta
X_{\bsk}\vert\,\sim\,1$, as envisaged here. Furthermore, since there
exists one resonant wavenumber $k_x$ for each $\bskp$, one should
expect that non-linear corrugation takes place on all inertial scales
present in the incoming turbulence spectrum. Note that these
wavenumbers are specified in the downstream rest frame; in the
upstream rest frame, which moves relative to the former with Lorentz
factor $\gamma_1$ and velocity $\beta_1$, $k_{x\vert
  1}\,=\,k_x/\gamma_1 - \beta_1\omega_{\rm A\vert 1}/c$,
i.e. $k_{x\vert1}\,\simeq\,-\beta_1\beta_{\rm A\vert1}k_z$ for
$\gamma_1\,\gg\,1$ and an Alfv\'en wave of frequency $\omega_{\rm
  A\vert 1}\,=\,\beta_{\rm A\vert1}k_z c$.

The sourcing of corrugation through MHD instabilities seeded in the
upstream plasma by the accelerated particles themselves represents an
interesting alternative.  Such instabilities have been discussed in
\citet{2009MNRAS.393..587P} and studied through dedicated numerical
simulations in \citet{2013MNRAS.433..940C}. They can be seen as a
generalization of the Bell instability \citep{2004MNRAS.353..550B} to
the relativistic regime and for perpendicular shocks: the existence of
a net charge or current of supra-thermal particles executing Fermi
orbits in the shock precursor then destabilizes magnetosonic modes of
the upstream plasma.  If magnetosonic waves are amplified, one may
then expect them to induce a resonant response of the shock. Of
course, the sourcing of corrugation by the accelerated particles
brings in an amusing chicken-and-egg problem if corrugation is a
necessary condition for the development of the relativistic Fermi
process, as advocated in the previous section.

\subsection{Application to the Crab nebula}
Let us finally discuss how the above discussion might apply to the
termination shock of the Crab pulsar wind. One should first point out,
however, that the existence of dissipation has been demonstrated by
\citet{2013MNRAS.428.2459K}, through the comparison between the
present content in magnetic energy in the Crab nebula and that input
over its lifetime.

In terms of spectral energy distribution, it is well-known that one
can reproduce the main observational features by assuming the
existence of a broken power-law of the pair population
\citep{1996MNRAS.278..525A}, with index $s\,\simeq\,1.6$ for ${\rm
  d}N/{\rm d}\gamma$ for $\gamma\,\lesssim\,\gamma_{\rm b}$, and
$s\,\simeq\,2.3$ above; the break Lorentz factor inferred is of order
$\gamma_{\rm b}\,\simeq\,2\times 10^6$. The maximum synchrotron photon
energy is about $100\,$MeV, corresponding to the radiation reaction
limit energy $\,\sim\,m_ec^2/\alpha_{\rm e.m.}$ ($\alpha_{\rm
  e.m.}\,\simeq\,1/137$ the electromagnetic fine structure
constant). Finally, the magnetic field inferred from a modelling of
the nebula, $B\,\sim\,200\,\mu$G, corresponds to rough equipartition
with the pair population.

In the present model, this rough equipartition is a natural
consequence of magnetic dissipation of turbulence into the pair
population. Furthermore, as argued in Sec.~\ref{sec:dissip}, the
pre-acceleration of particles in the turbulence seeded by corrugation
may also produce a power-law with an index $s$ comprised between $1$
and $2$, as observed. If ${\rm d}N/{\rm
  d}\gamma\,\propto\,\gamma^{-s}$ with $1<s<2$ below $\gamma_{\rm b}$,
then the break Lorentz factor
\begin{equation}
  \gamma_{\rm b}\,\simeq\,\gamma_1\,\left(\frac{\gamma_{\rm
      d}}{\gamma_1}\right)^{\frac{1}{2-s}}
\end{equation}
where
\begin{equation}
  \gamma_{\rm d}\,\simeq\,\epsilon_e\,\frac{L_{\rm w}}{\dot N m_e c^2}
\end{equation}
is the mean Lorentz factor per particle, after a fraction $\epsilon_e$
of the wind luminosity $L_{\rm w}$ has been transferred in the $\dot
N$ pairs advected through the shock per unit time. Assuming a
multiplicity $\kappa\,=\,10^6\kappa_6$ beyond the standard
Goldreich-Julian injection rate $\dot N_{\rm GJ}\,\simeq\,
e^{-1}\sqrt{L_{\rm w}c}$~\citep{1969ApJ...157..869G}, as seems
required for the Crab, one finds $\gamma_{\rm d}\,\simeq\,5\times
10^4\kappa_6^{-1}$ for $\epsilon_e\,\sim\,1$. Setting $\gamma_{\rm
  b}\,\simeq\,2\times 10^6$ for $s\,=\,1.6$ then requires a wind
Lorentz factor
\begin{equation}
  \gamma_1\,\sim\, 4\times 10^3\kappa_6^{-1.7}\ .
\end{equation}
From a theoretical point of view, this value seems appealing, because
it somewhat alleviates the requirements regarding the acceleration of
the wind, which represents a nagging issue in this field
\citep[e.g.][]{2009ASSL..357..421K}.

Detailed numerical simulations of the Crab nebula have shown that it
is possible to explain the main morphological features of this nebula
provided the shock accelerates pairs efficiently \citep[e.g., for a
  review][]{2015SSRv..191..391K}. Interestingly, the nebula reveals
slow moving structures called ``wisps'', originating from the
termination shock; those features have a typical angular size $1''$,
corresponding to roughly $0.01\,{\rm
  pc}\,\sim\,0.1R$~\citep[e.g.][]{2013MNRAS.433.3325S}. If those wisps
are interpreted as long-lived modes produced by the corrugated shock,
then it argues in favor of corrugation up to scales close to
$0.1R$. Since the highest energy pairs in the nebula have an energy
$\sim\,1\,$PeV, the maximum gyro-radius of accelerated particles is
$r_{\rm g,max}\,\sim\,0.01\,$pc for a nebular field of $100\,\mu$G,
i.e. of the same order of magnitude as the size of the wisps. In the
context of the above discussion, which has suggested that the
accelerated particles could seed corrugation on scales $r_{\rm g}$
through instabilities in the shock precursor, this connection is
rather intriguing. Moreover, the possibility of corrugation up to
scales $k^{-1}\,\sim\, r_{\rm g,max}$ indicates that Fermi
acceleration should be operative up to those scales. As discussed in
Sec.~\ref{sec:Fermi}, acceleration should further proceed with an
acceleration timescale $\sim\,r_{\rm g}/c$, leading to Bohm type
acceleration up to the radiation reaction limit.

In this regard, one may note that other models trying to explain the
pre-acceleration of pairs in the nebula and the dissipation through
magnetic reconnection generally fail to account for Bohm acceleration
to high energies, because the annihilation of the magnetic field in
the striped part of the wind leaves behind a short scale turbulence,
with a slow scattering timescale, hence leading to a maximal energy
well below that observed in the Crab nebula, see e.g.
\citet{2013ApJ...771...54S}.

\section{Conclusions}\label{sec:conc}
The present paper has speculated that the termination shock of pulsar
winds might be strongly corrugated. It has discussed possible sources
of corrugation and exhibited various interesting phenomenological
consequences.

Corrugation could in principle be excited by the interaction of
downstream fast magnetosonic modes catching up the shock front,
through the advection of upstream turbulence modes, or through the
generation in the shock precursor of turbulence by particle
acceleration. As discussed here and in \citet{LRG16}, the latter two
possibilities are more interesting in the present context because of
the existence of a resonance in the response of the shock to upstream
perturbations, leading to possible large amplification of turbulent
modes.

Once corrugation is excited on a range of scales, a fraction of order
unity of the incoming ordered magnetic energy is converted into
turbulence, immediately downstream of the shock. This conversion slows
down appreciably the flow velocity along the shock normal, which could
help understand why the post-shock nebula moves so slowly in the
pulsar rest frame, in accord with the seminal discussion of
\citet{1984ApJ...283..694K}. Various dissipative effects could then
transfer a sizable fraction of the turbulence energy density into the
pair population. In particular, slow magnetosonic modes are rapidly
dissipated in a relativistic plasma. A direct consequence is that,
independently of the upstream magnetization of the flow, the
downstream magnetization beyond this dissipative layer would decrease
to values of order unity. This, of course, has significant virtues for
understanding the phenomenological properties of the Crab nebula,
which indeed reveals a rough equipartition between the pairs and the
magnetic energy content.

The pre-acceleration of the pairs in the dissipative layer through
stochastic acceleration leads to the emergence of a power-law, with an
index $s$ typically between $1$ and $2$, because stochastic
acceleration is balanced by escape losses due to advection outside of
the dissipative layer. The present paper has also speculated that the
excitation of turbulence on a broad range of scales behind the shock
could sustain a relativistic Fermi process with a Bohm-type
acceleration timescale; this point remains to be demonstrated however,
using for instance dedicated test-particle simulations. It has then
been shown that this combination of stochastic pre-acceleration
followed by Fermi acceleration could potentially help understand the
spectral features of the Crab nebula, provided the Lorentz factor of
the termination shock is $\gamma_1\,\sim\,4\times 10^3$ in the nebula
rest frame (assuming a pair multiplicity $\kappa\,\sim\,10^6$).

Further work is required along several lines to test this speculative
model. In particular, dedicated numerical simulations are needed to
understand the physics of corrugation in the non-linear regime through
the interaction of a relativistic magnetized shock with upstream
perturbations. As mentioned above, dedicated numerical simulations
would also be needed to understand how such a corrugated shock can
accelerate particles, and with what efficiency. It would be
interesting to understand how the accelerated particles could
themselves seed perturbations in the upstream plasma, and how such
perturbations could influence the shock corrugation pattern. Finally,
such simulations would have to be properly placed in a global context
to understand the impact of the nebular turbulence on the shock
itself.\bigskip

{\bf Acknowledgments:} it is a pleasure to thank A. Bykov,
L. Gremillet, R. Keppens, G. Pelletier and O. Ramos for insightful
discussions and advice. This work has been financially supported by
the Programme National Hautes \'Energies (PNHE) of the C.N.R.S. and by
the ANR-14-CE33-0019 MACH project.

\bibliographystyle{jpp}

\bibliography{shock}

\end{document}